\documentclass[usenatbib]{basi}
\usepackage{rotating}
\usepackage{dcolumn}
\usepackage{graphicx}
\usepackage{graphics}
\usepackage{longtable}

\begin{document}

\title[Near IR evolution of V476 Sct]{Near-infrared spectroscopic and photometric evolution of Nova V476 Scuti  - a nova that formed optically thin dust}

\author[R. K. Das et~al.]%
       {R. K. Das$^{1}$\thanks{email: \texttt{ramkrishna.das@bose.res.in}},
       D. P. K. Banerjee$^{2}$, N. M. Ashok$^{2}$ and Soumen Mondal$^{1}$ \\
       $^1$Astrophysics and Cosmology Department, S N Bose National Centre for Basic Sciences,\\
       Salt Lake, Kolkata 700 098, India\\
       $^2$Astronomy and Astrophysics Division, Physical Research Laboratory, Ahmedabad - 380 009, \\ Gujarat, India}

\pubyear{2013}
\volume{41}
\pagerange{\pageref{firstpage}--\pageref{lastpage}}

\date{Received 2013 March 07; accepted 2013 July 12}

\maketitle

\label{firstpage}

\begin{abstract}
We present results of near-infrared (near-IR) $JHK$ (1.07 - 2.5 $\mu$m) spectroscopic and
photometric observations of Nova V476 Scuti (V476 Sct) which was discovered in outburst
in 2005 September. The near-IR observations of the nova presents the evolution of the
post-maxima spectra and near-IR light curve. The spectra of V476 Sct, observed on 9 different
epochs, show prominent lines due to  HI, OI, CI and NI. Based on the IR spectral signatures
we independently identify it as a Fe II type of nova, consistent with the same classification
obtained from optical spectra. A detailed identification of the observed spectral lines
is presented. The near-IR $JHK$ light curve extending for a period of about 59 days after
outburst clearly shows the formation of a optically thin dust shell, a phenomenon
which is not commonly observed in novae.  By fitting black body curves to the spectral
energy distributions (SEDs) the temperatures of the dust shell on different epochs have
been estimated. Dust formation in V476 Sct is consistent with the presence of lines
of elements with low-ionization potential like Na and Mg in the early spectra
which had  earlier been suggested by us to be potential indicators of dust formation
at a later phase  in a nova's development.
\end{abstract}

\begin{keywords}
infrared: stars -- line: identification -- stars: novae, cataclysmic variables -- stars: individual: V476 Sct
-- techniques: spectroscopic -- techniques: photometric
\end{keywords}

\section{Introduction}

Nova V476 Scuti (V476 Sct) was discovered in outburst towards the end of September 2005 independently by
several observers. Takao (2005) reported its presence (mag 10.3) on an unfiltered CCD image taken on September
30.522 UT. Yamaoka (2005), reported that  the new object appeared at a magnitude of about 10
on ASAS-3 images taken on September 28.088 and October 1.025, but was not present on
September 24.629 (limiting mag perhaps 13).  This implies that the object was caught early
in its outburst. Independent discoveries of the object at magnitude about 10.9 was also made by Haseda (2005)
from photographs taken on September 30.417 and 30.422 and also by Gilmore \& Kilmartin (2005)
from short unfiltered CCD exposures taken on October 1.393
showing the new object at mag 10.0. However, for the present purpose we assume that September 30.522 UT
is the outburst day. Gilmore \& Kilmartin (2005)
provide precise coordinates for the nova
($\alpha$ = 18$^{h}$ 32$^{m}$ 04.75$^{s}$, $\delta$ = - 0.6$^{\circ}$ 43$^{\prime}$ 34$^{\prime\prime}$) with
an accuracy of 0.1$^\prime$ and show that a comparison with a Digitized Sky
Survey (DSS) red plate from 1988 August 10  presents no obvious precursor at the nova's
position though a star of red mag 17.9 lies slightly to the east of the nova.

A low-resolution spectrogram (range 550-1050 nm) was obtained  on October 6.36 UT by Kiss et al. (2005), showing
that V476 Sct had a spectrum typical of a classical nova. Strong emission lines of
hydrogen, oxygen, calcium, magnesium, carbon, and iron were
identified with H$\alpha$  having a symmetric profile with FWZI exceeding 4000 km s$^{-1}$.  The strongest emission
lines seen include H$\alpha$, the O{\sc i} 777.3, 844.6, and 926.4 nm lines,
members of the hydrogen Paschen series, the infrared (IR) Ca triplet,
and the C{\sc i} 940.6 nm line, all with FWZI greater than 2000 km s$^{-1}$. The authors remark that the overall
spectral appearance is similar to that of the Fe II nova  V443 Sct (N Sct 1989),
two weeks after its maximum (Williams et al. 1991).
CCD photometry and spectroscopy were also done by Munari et al. (2006) who showed
it to be a fast nova, characterized by $t_{2}$ = 15 and $t_{3}$ = 28 d and affected by an E(B-V) $\sim$ 1.9 magnitude  reddening. From an MMRD (maximum magnitude vs. rate of decline) analysis they estimate the distance to the nova to be is 4$\pm$1
kpc, and its height above the Galactic plane to be z = 80$\pm$20 pc. They classified the nova as belonging to the Fe II type.
The classification is consistent with the theoretical result obtained by Hachisu $\&$ Kato (2007). Their model
light curve generated for CO nova 2 ($X$ = 0.35, $X_{CNO} = 0.30$, $Z$ = 0.02) nicely fitted with the observed visual and V magnitudes.
From this model calculation the mass of white dwarf in nova V476 Sct was estimated to be 0.95$\pm$0.05 $M_{\odot}$.

The only IR study of the object was done by Perry et al. (2005) who obtained 0.47- to 2.5-$\mu$m spectroscopy of V476
Sct on November 15.108 UT.  The nova was still quite early in its spectral
development, showing structured emission in the lines of C{\sc i}, N{\sc i},
and Fe{\sc{ii}} with FWHM of 1600 km s$^{-1}$.  He{\sc i} lines were just emerging and
were still very weak.  The strongest lines in the visible and
IR spectrum were from O{\sc i}, and these are produced almost
completely by Lyman $\beta$ fluorescence. The authors estimate the
reddening as indicated by the O{\sc i} lines to be E(B-V) = 2.0. In the X-ray regime, no super
soft X-ray phase was detected from the source and only an upper limit could be placed (Ness et al. 2007).

In this study we present  near-infrared (near-IR) spectra of the object taken during the early decline
and analyze their general properties.
We also present near-IR photometric coverage of the object that shows the development
of a dust shell around the nova. Although novae are known to form dust,  the distinguishing characteristic of
the dust shell in V476 Sct is that it is optically thin. The present study presents an example,  not too often
encountered in the literature, of the IR development of such an optically thin dust shell.

%************************************* Table 1 **************************************************
\begin{table}
 \centering
\caption{Log of the photometric observations of V476 Sct. The date of outburst
is taken to be 2005 September 30.522 UT.}

\begin{tabular}{@{}lcccccccc@{}}

%\hline
\hline
Date UT    &  Days since    &              &         Magnitudes          &          \\
(2005)   &  outburst  &      $J$         &          $H$          &     $K$     \\
%\hline
\hline
06.664 Oct	&	6.14	&	7.24	$\pm$	0.06	&	6.53	 $\pm$	0.04    &   6.02	 $\pm$	 0.12\\	
11.615 Oct	&	11.09	&	7.80	$\pm$	0.09	&	7.06	 $\pm$	0.09	&	6.77	 $\pm$	0.12	 \\
12.611 Oct	&	12.09	&	7.85	$\pm$	0.07	&	7.35	 $\pm$	0.04	&   7.05	 $\pm$	0.10	 \\
15.692 Oct	&	15.17	&	8.06	$\pm$	0.06	&	7.78	 $\pm$	0.16	&	7.34	 $\pm$	0.50	 \\
16.688 Oct	&	16.17	&	8.24	$\pm$	0.08	&	7.73	 $\pm$	0.11	&	6.82	 $\pm$	1.14	 \\
20.634 Oct	&	20.11	&	8.28	$\pm$	0.11	&	7.07	 $\pm$	0.03	&	6.18	 $\pm$	0.06	 \\
25.645 Oct	&	25.12	&	8.37	$\pm$	0.06	&	7.09	 $\pm$	0.07	&	5.75	 $\pm$	0.13	 \\
26.587 Oct	&	26.07	&	8.29	$\pm$	0.04	&	7.01	 $\pm$	0.03	&	5.69	 $\pm$	0.04	 \\
07.572 Nov	&	38.05	&	8.78	$\pm$	0.06	&	6.72	 $\pm$	0.08	&	5.02	 $\pm$	0.03	 \\
08.607 Nov	&	39.09	&	8.64	$\pm$	0.19	&	6.51	 $\pm$	0.02	&	5.02	 $\pm$	0.06	 \\
14.588 Nov	&	45.07	&	9.20	$\pm$	0.09	&	6.98	 $\pm$	0.04	&	5.25	 $\pm$	0.04	 \\
17.592 Nov	&	48.07	&	9.52	$\pm$	0.03	&	7.18	 $\pm$	0.04	&	5.40	 $\pm$	0.02	 \\
24.572 Nov	&	55.05	&	10.10	$\pm$	0.05	&	7.59	 $\pm$	0.08	&	5.66	 $\pm$	0.08	 \\
01.564 Dec	&	62.04	&	11.39	$\pm$	0.52	&	8.70	 $\pm$	0.25	&	6.15	 $\pm$	0.14	 \\
02.562 Dec	&	63.04	&	11.54	$\pm$	0.35	&	8.70	 $\pm$	0.14	&	6.33	 $\pm$	0.16	 \\

\hline
\end{tabular}
\end{table}
%************************************* Table 1 **************************************************

%*******************************Table 2**************************************
\begin{table}
\begin{center}
%{\bf\large{Table 1}} \\~\\
\caption{Log of the spectroscopic observations of V476 Sct. The date of outburst
is taken to be 2005 September 30.522 UT.}

\begin{tabular}{lcccccc}
%\hline
\hline
Date & Days        & &         & Integration time &   \\
2005 & since        & &        & (sec)            &    \\
(UT) & outburst  & & \emph{J} & \emph{H}         & \emph{K}    \\
\hline
%\hline

Oct 5.666 & 5.876   & & 40      & 40               & 60 \\
Oct 6.618 & 6.826   & & 120      & 120               & 120 \\
Oct 11.669 & 11.877   & & 120      & 120               & 120 \\
Oct 15.610 & 12.798   & & 180      & 180               & 180 \\
Oct 16.623 & 15.820   & & 60      & 45               & 60 \\
Oct 19.621 & 16.832   & & 60      & 60               & 60 \\
Oct 21.618 & 19.831   & & 200     & 200               & 200 \\
Oct 25.593 & 21.830   & & 300     & 300               & 300 \\
Nov 03.591 & 34.802   & & 300     & 300               & 300 \\

\hline
\end{tabular}
\end{center}
\end{table}
%*******************************Table 2**************************************

\section{Near-IR Mount Abu observations}
The near-IR spectroscopic and photometric observations of V476 Sct were carried out from
1.2m Mount  Abu telescope, operated by the Physical Research Laboratory.
The $JHK$ spectra  were
obtained at similar dispersions of $\sim$ 9.5 Angstrom/pixel in each of the
$J, H, K$ bands using the Near-Infrared Imager/Spectrometer which uses a
256$\times$256 HgCdTe NICMOS3 array. A set of two spectra were
taken with the object dithered to two positions along the slit which were subtracted from each other to
eliminate the sky contribution and the detector dark counts. The
spectra were then extracted using APEXTRACT package in IRAF and wavelength calibration was
done using a combination of OH sky lines and telluric lines that
register with the stellar spectra. Following the standard procedure,
the object spectra were then ratioed with the spectra of a comparison
star (SAO 143021; spectral type B9V; $V$ = 3.427) observed at similar airmass as the object and from whose spectra
the Hydrogen Paschen and Brackett absorption lines had been removed. The ratioed spectra were then multiplied by a blackbody curve at the effective temperature of the comparison star to yield the final spectra. Though the ratioing process removes the telluric features
sufficiently well, some residuals are still left in the wavelength regions where
telluric absorption is strong. This applies significantly to the  regions around 1.12 $\mu$m in the $J$ band
and between 2 to 2.05 ${\rm{\mu}}$m in the $K$ band. Hence care should be taken while interpreting any feature appearing in these regions.\\

Photometry in the $JHK$ bands was done under photometric sky conditions using
the imaging mode of the NICMOS3 array. Several frames, at five
dithered positions offset typically by 20 arcsec, were obtained of
both the nova and a selected standard star (SAO 162177) in each of the $J, H, K$
filters. Near-IR $JHK$ magnitudes were then derived using APPHOT package in IRAF tasks
following the regular procedure followed by us for photometric
reduction (e.g. Banerjee \& Ashok 2002; Naik et al. 2010).
The log of the photometric and spectroscopic observations are given in Tables 1 and 2 respectively.
 We could not extend our observations beyond early December because the object had at that stage begun to  approach conjunction with the Sun.

%*******************************Figure 1**************************************
\begin{figure}
\centering
\includegraphics[bb= 0 0 693 493, width=5.3 in, height = 4.7 in, clip]{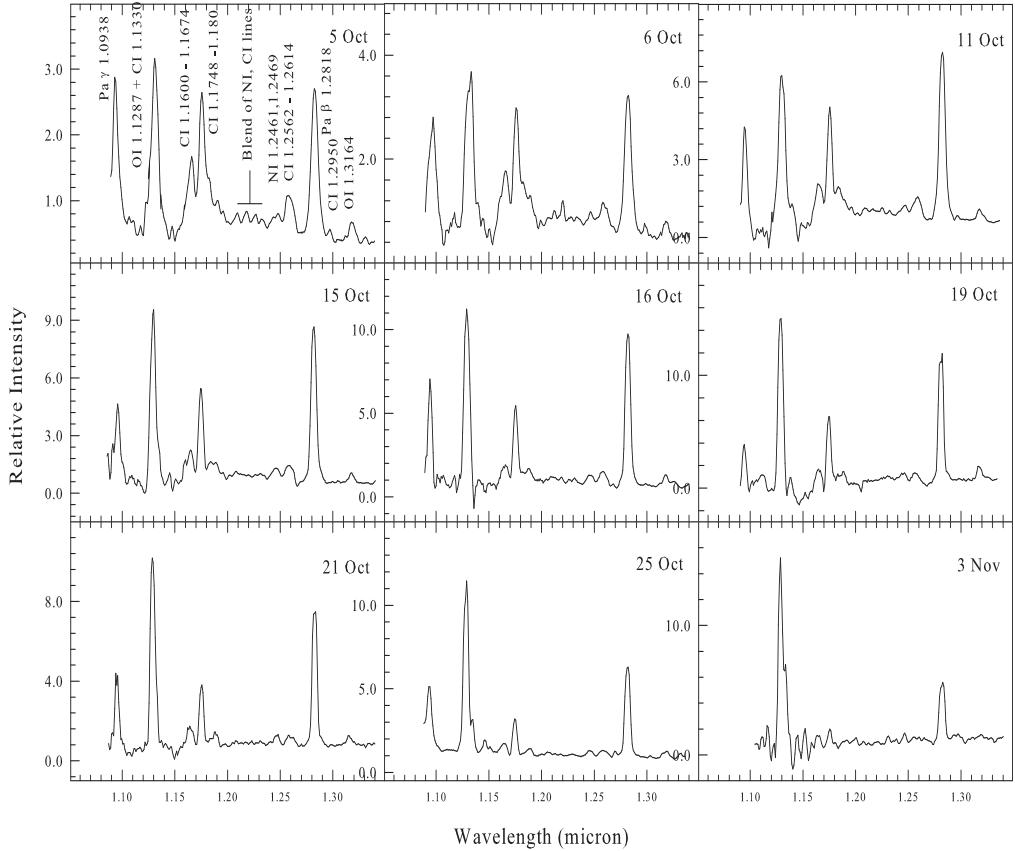}
\caption{$J$ band spectra of V476 Sct with the prominent lines marked.  All spectra are normalized to unity at 1.25 $\mu$m.
  }
\end{figure}
%*******************************Figure 1**************************************

%*******************************Figure 2**************************************
\begin{figure}
\centering
\includegraphics[bb= 0 0 693 486, width=5.3 in, height = 4.7 in, clip]{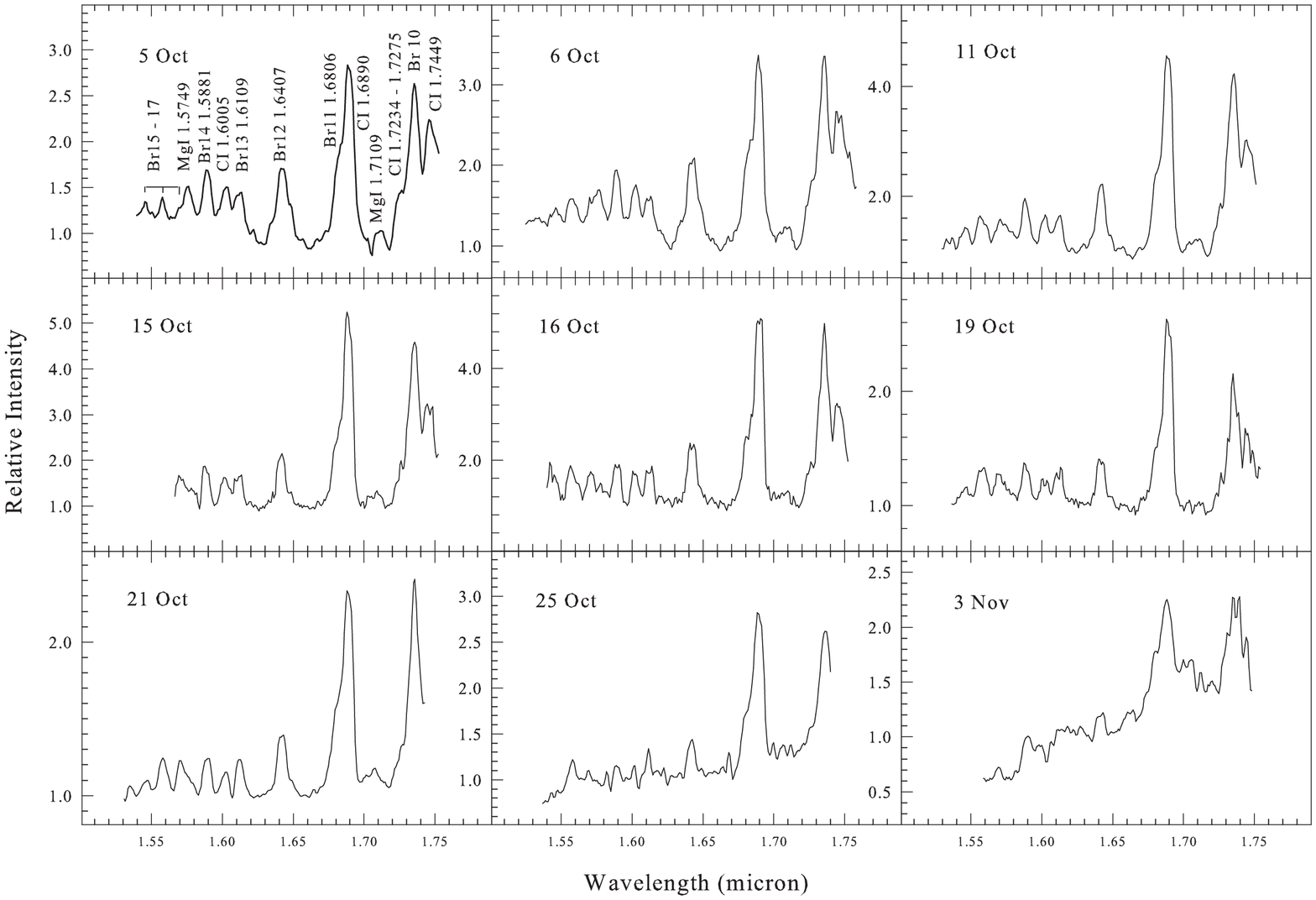}
\caption{The $H$ band spectra of V476 Sct with prominent lines marked (see Table 3 for details). All spectra are normalized to unity at 1.65 $\mu$m.
  }
\end{figure}
%*******************************Figure 2**************************************

%*******************************Figure 3**************************************
\begin{figure}
\centering
\includegraphics[bb= 0 0 693 486, width=5.3 in, height = 4.7 in, clip]{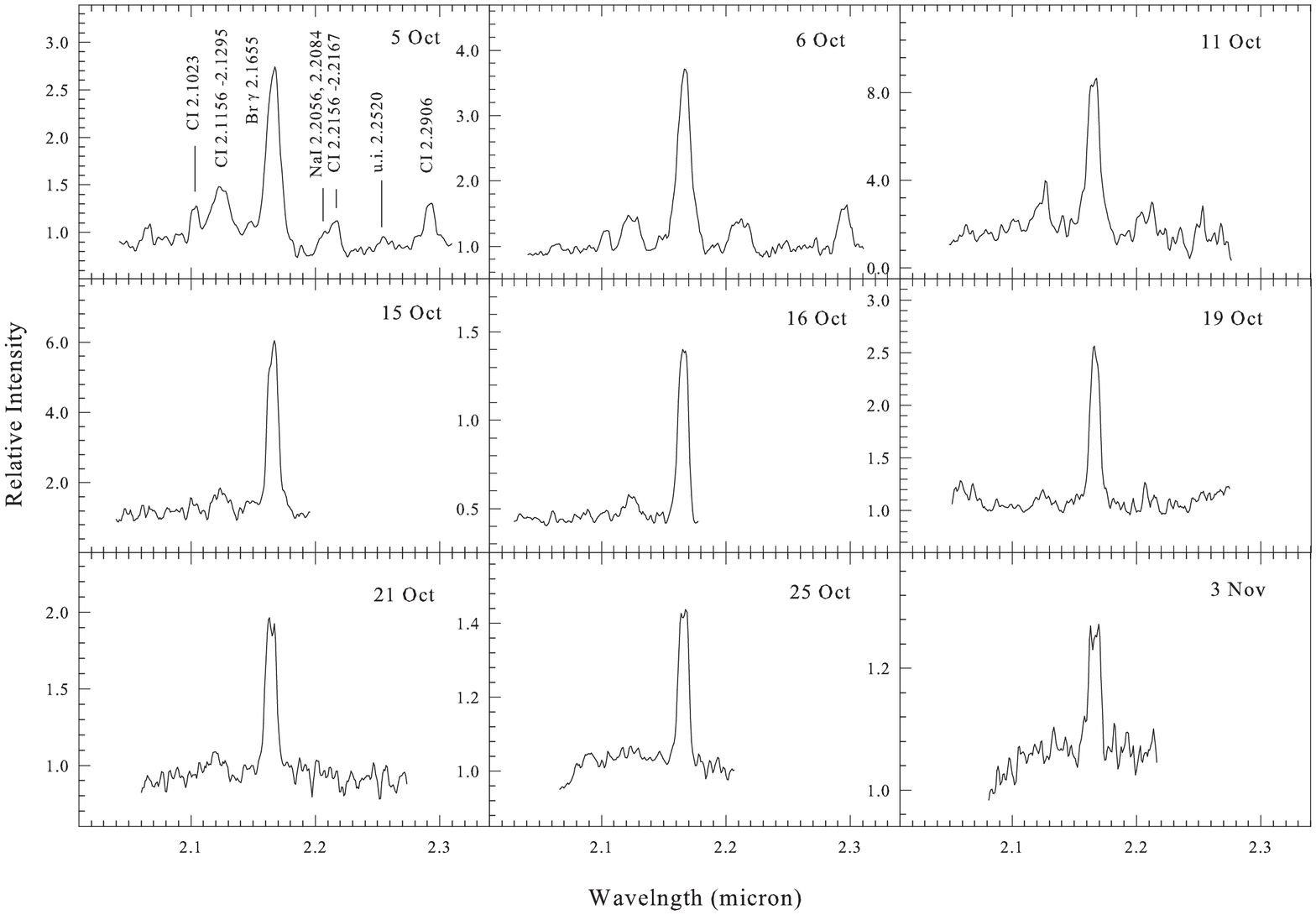}
\caption{$K$ band spectra of V476 Sct with prominent lines marked (see Table 3 for details).
 All spectra are normalized to unity at 2.2 $\mu$m.
}
\end{figure}
%*******************************Figure 3**************************************

\section{Results}
\subsection{General characteristics of the near-IR spectra}
The mosaics of $J$, $H$ and $K$ band spectra of V476 Sct observed on 9 different epochs
are presented in Figs. 1, 2 and 3 respectively. Prominent lines are marked on the Figures.
In addition to the strong lines due to H{\sc i}, O{\sc i}, C{\sc i} and N{\sc i} as marked on Figs. 1, 2 and 3, weaker
features may also be seen when the spectra are magnified. A detailed list of the lines is presented in Table 3.
The early spectra of V476 Sct  display the typical IR signatures  which are expected  of a FeII nova.
In addition to lines of H, N and O, these novae show strong lines of carbon e.g. in the $J$ band in
the 1.16 to 1.18 $\mu$m wavelength region and in the $H$ band in the spectral region around Br 10
at 1.73 $\mu$m  and further redwards. These C{\sc i}
lines are strongly seen in the spectra presented here. In contrast, such carbon lines are absent or very weak
in the spectra of He/N novae e.g. in the novae V597 Pup and V2491 Cyg (Naik et al. 2009). A detailed paper showing the
IR classification of novae spectra and the differences between FeII and He/N novae is given in Banerjee \& Ashok (2012). The
FeII nova classification, independently arrived at from the IR spectra,  is consistent with  its optical classification
as a FeII type nova. Examples of similar IR spectra of FeII novae   may be seen in V1280 Sco (Das et al. 2008) and
V2615 Oph (Das et al. 2009),  V2274 Cyg (Rudy et al. 2003), V1419 Aql (Lynch et al. 1995) and  V5579 Sgr (Raj et al. 2011)

The most prominent lines in the $JHK$ spectra of V476 Sct are the Paschen and Brackett hydrogen recombination lines.
The average FWHM (full width at half maxima) of Pa $\beta$ and Br $\gamma$ are respectively 1529 km s$^{-1}$ and 1425 km s$^{-1}$ which
yields the expansion velocity $\sim$738 km s$^{-1}$. This result is in line with the value of $\sim$600 km s$^{-1}$
estimated from the optical data (Munari et al. 2006). The O{\sc i} line at 1.1287 is very strong compared to the
continuum fluoresced O{\sc i} 1.3164 $\mu$m indicating that Lyman $\beta$ fluorescence significantly contributes to
the strength of O{\sc i} 1.1287. We do not see signs of any He{\sc i} lines which is consistent with the report of Perry
et al. (2005) that He{\sc i} lines were just emerging and were still very weak even on 2005 November 15.

The region between 1.2 to 1.275 ${\rm{\mu}}$m contains the complex blend of a  large number of  N{\sc i} and C{\sc i} lines that
are  often seen in the early spectra of FeII novae. The presence of lines of Na{\sc i}  at 2.2056
 and 2.2084 $\mu$m  are worth noting as it may be inferred, from their presence, that dust will form in a nova.
During the analysis of V1280 Sco, one of the conclusions  that emerged, was that the  lines of Na{\sc i} particularly
are associated with low excitation and ionization conditions (Das et al. 2008). Such conditions necessarily imply
the  existence  of a cool zone which  is conducive for dust formation. Such regions could be  associated with
clumpiness in the nova ejecta which likely provide the cool dense sites needed for molecule and dust formation.
From our earlier studies (Banerjee $\&$ Ashok, 2012) we have observed  that  whenever these lines are seen,
dust did indeed form in the nova.  Examples of such  novae which showed these lines and simultaneously formed dust are V2274 Cyg (Rudy et al. 2003), V1419 Aql (Lynch. et al. 1995),  NQ Vul
(Ferland et al. 1979), V705 Cas (Evans et al. 1996), V842 Cen (Wichmann et al. 1991),
V1280 Sco (Das et al. 2008),  V5579 Sgr (Raj et al. 2011) and V496 Sct (Raj et al. 2012).
The presence of these lines in V476 Sct, and the dust formation witnessed subsequently,
is consistent with this scenario.
In addition to the presence of the Na and Mg lines, it is possible that there are other
parameters which influence and correlate with dust formation (see for e.g. Gehrz 1988; Bode \& Evans 2008 and references therein).
However a detailed discussion or review of all such parameters  is beyond the scope of this work.
We are also unable to comment whether CO formed in this nova,
as it did for e.g. in V2615 Oph (Das et al. 2009) and V496 Sct (Raj et al. 2012 and
references therein),  because our $K$ band spectra do not extend much beyond 2.29 $\mu$m  where
the CO first overtone bands are expected.

%*******************************Table 3**************************************
\begin{table}
\begin{center}
%{\bf\large{Table 1}} \\~\\
\caption{List of observed lines in the $JHK$ spectra of V476 Sct.}
\begin{tabular}{lllllll}

%\hline
\hline
Wavelength      & Species       & Other contributory & Wavelength   & Species           & Other contributory \\
($\mu$m)        &               & species/ remarks   & ($\mu$m)     &                   & species/ remarks\\

\hline
%\hline

1.0938   	    & Pa $\gamma$   &        	        & 1.5749             & Mg{\sc i}  & Mg{\sc i}  1.5741, 1.5766,  \\
1.1190          & Na{\sc i}      &                   &                    &                   & \& C{\sc i}  1.5784   				 \\
1.1287	        & O{\sc i}       &                   & 1.5881             & Br 14             & C{\sc i} 1.5853 				 \\
1.1330          & C{\sc i}       &                   & 1.6005   		& C{\sc i}   & \\
1.1381          & Na{\sc i}      & C{\sc i}  1.1373        & 1.6109   		& Br 13             & \\
1.1404          & Na{\sc i}      & C{\sc i}  1.1415        & 1.6335 	    & C{\sc i}	&  \\
1.1600-1.1674   & C{\sc i}       & strongest lines	& 1.6419   		& C{\sc i}   &   \\
   		        &               & at 1.1653, 1.1659,& 1.6407  		& Br 12             & C{\sc i} lines between	\\
                &               & 1.16696           & 1.6806        & Br 11             & 1.6335-1.6505 	 \\
1.1748-1.1800   & C{\sc i}       & strongest lines   & 1.690        & C{\sc i}    &   \\
   		        &               & at 1.1748, 1.1753,& 1.7045       & C{\sc i}    & \\
   		        &               & 1.1755            & 1.7109   	   & Mg{\sc i}   &     \\
1.1828          & Mg{\sc i}      &                   & 1.7234 -      & C{\sc i}    & Several C{\sc i}  lines  \\
1.1819-1.1896 	& C{\sc i}       & strongest lines at & 1.7275    		        &    	             & in this region  \\
                &               & 1.1880, 1.1896     & 1.7362  		    & Br 10  	& C{\sc i}  1.7339    \\
1.1949          & u.i           &                    & 1.7449             & C{\sc i}  & 		 \\
1.2074,1.2095   & N{\sc i}       & C{\sc i}  1.2088         & 2.1023             & C{\sc i}              &  \\
1.2187,1.2204  	& N{\sc i}       &              & 2.1156 -            & C{\sc i}    & strongest lines at \\
1.2249,1.2264   & C{\sc i}       &              & 2.1295                   &			      & 2.1156, 2.1191,  	\\
1.2329          & N{\sc i}       &              &                          &               & 2.1211, 2.1260, \\
1.2382 		    & N{\sc i}       &              &  				          &               & and 2.1295\\
1.2461,1.2469   & N{\sc i}       & O{\sc i}  1.2464   & 2.1452                   & Na{\sc i}    & \\
1.2562,1.2569  	& C{\sc i}       & O{\sc i}  1.2570   & 2.1655                   & Br $\gamma$   	 &       	 \\
1.2601,1.2614   & C{\sc i}       &              & 2.2056		   & Na{\sc i}&  	                      \\
1.2818          & Pa $\beta$    &              & 2.2084 	   & Na{\sc i} & \\
1.2950        	& C{\sc i}       &              & 2.2156 - &            		& C{\sc i} 2.2156, \\
1.3164         	& O{\sc i}       &              & 2.2167			    &			        & 2.2160, 2.2167 			 \\
1.5439   	   	& Br 17         & 	           & 2.2520   	& u.i 			    &				 				 \\
1.5557   		& Br 16        	&   		   & 2.2906   	& C{\sc i}  	& \\
1.5685  		& Br 15        	&              &  &   &\\

\hline
\end{tabular}
\end{center}
\end{table}
%*******************************Table 3**************************************

%*******************************Figure 4**************************************
\begin{figure}
\centerline{\includegraphics[bb=0 418 333 751,width=4.0in, height=3.7 in,clip]{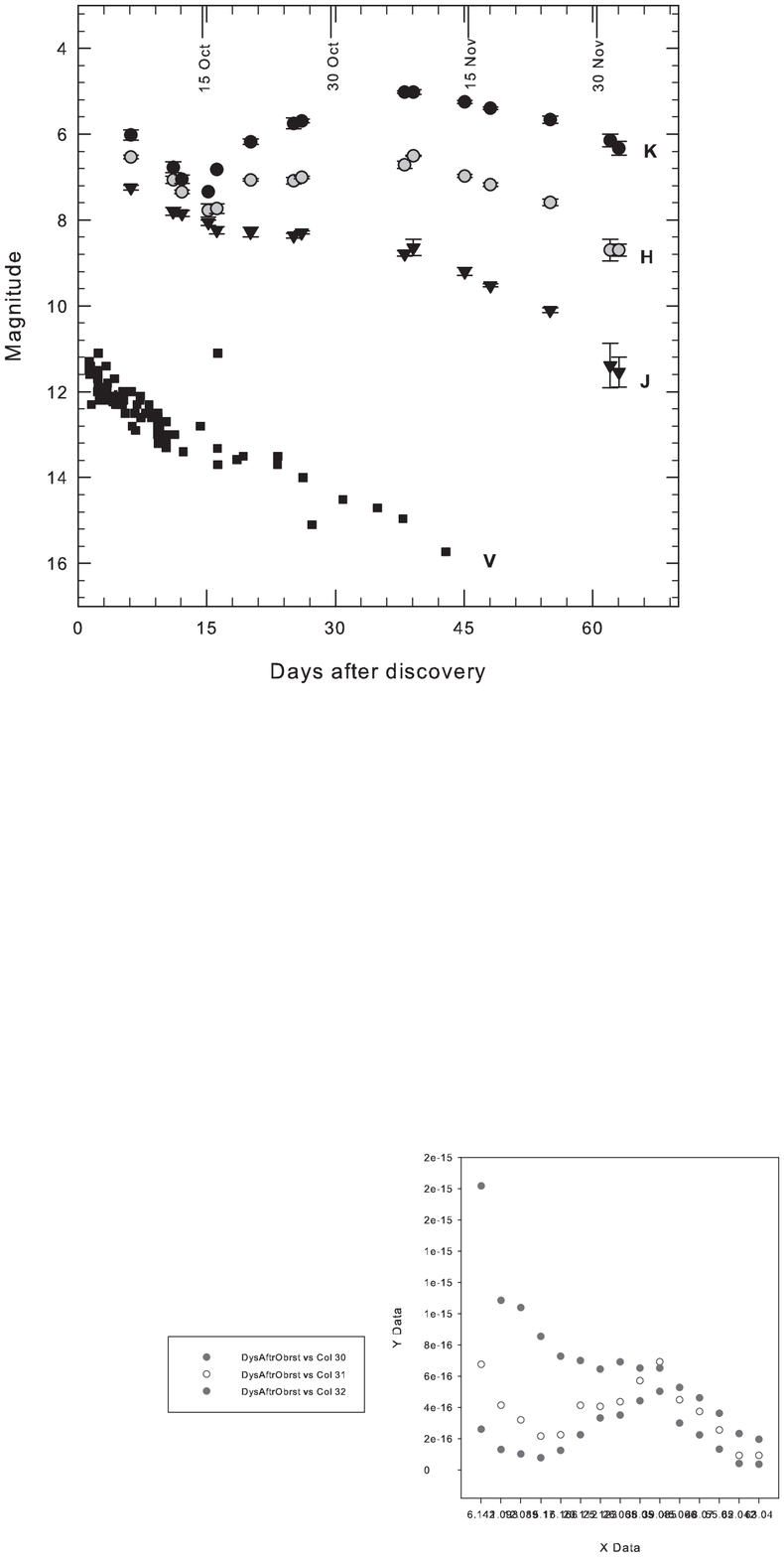}}
\caption[]{ The near-IR lightcurve in the $J, H $ and $K$ bands (filled black circles, gray circles and inverted triangles respectively)
during the early evolution of nova  V476 Sct which was discovered in outburst on 2005 September 30.522. The $V$ band curve,
based on AAVSO and AFOEV data,  is also shown. The near-IR magnitudes begin to increase around 15 - 16 October indicating
an IR excess due to dust formation. However, no sharp associated drop is seen in the optical lightcurve at this
stage indicating that the dust formed is optically thin.}
\label{fig2}
\end{figure}
%*******************************Figure 4**************************************

%*******************************Figure 5**************************************
\begin{figure}
\includegraphics[bb=  0 0 780 352, width=5.3 in, height = 4.0 in, clip]{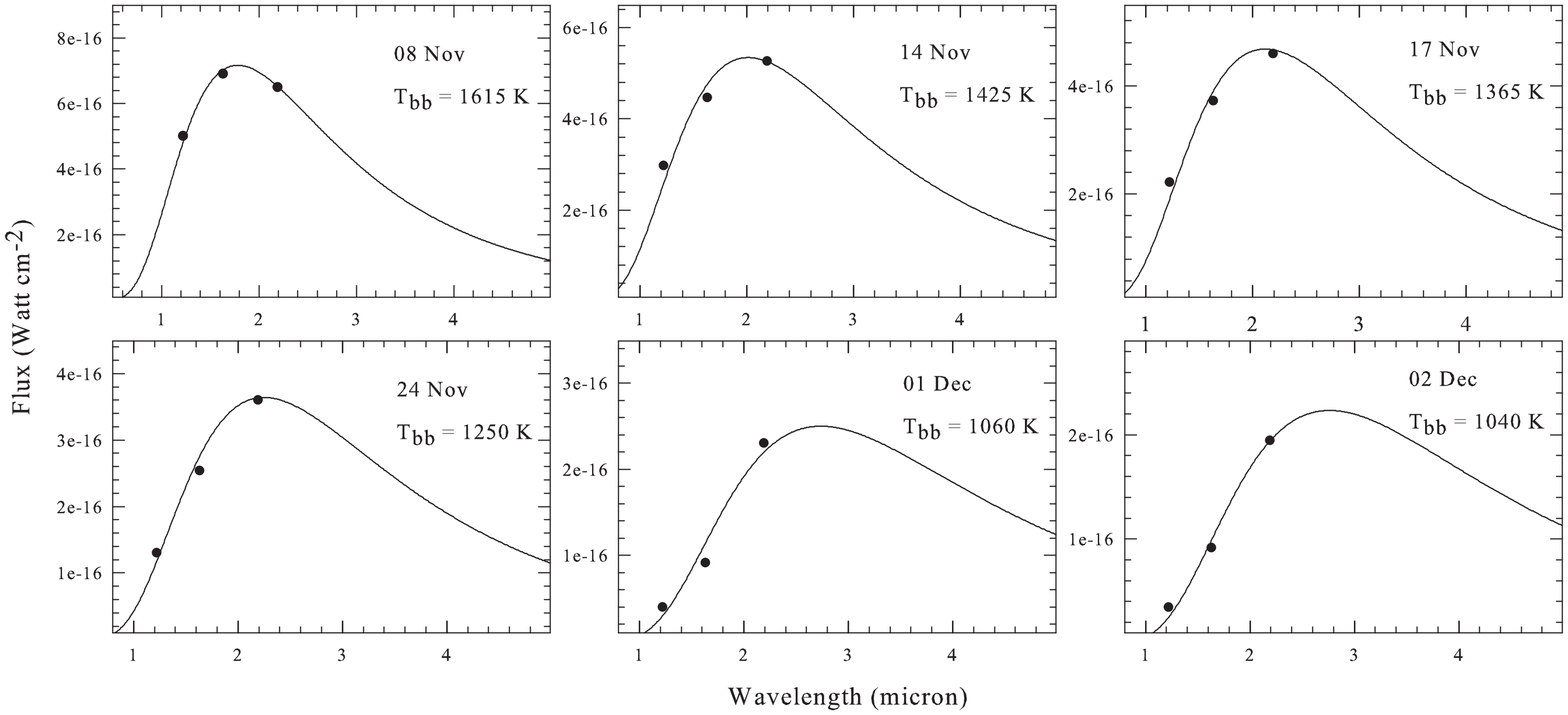}
\caption{Blackbody fits to the dereddened $JHK$ fluxes of V476 Sct at six epochs, beginning from 8 November 2005,  showing the cooling of the dust as the peak of the SED shifts towards longer wavelengths with time. For details see Section 3.2.}
\end{figure}
%*******************************Figure 5**************************************

\subsection{Dust formation}

The near-IR $JHK$ lightcurve along with  the optical lightcurve of V476 Sct is presented in
Fig. 4. The near-IR observations were made using Mt Abu telescope and the optical V band data are from AAVSO and AFOEV. The
near-IR $JHK$ magnitudes were corrected for extinction using Koornneef's (1983) relations and A$_{V}$ = 5.89 mag.
The value of A$_{V}$ was calculated assuming E(B - V) = 1.9 as estimated by Munari et al. (2006).
Our IR observations show a steady decline in brightness till October 15 ($\sim$ 15 days after discovery). At this stage each of the $J,H,K$ magnitudes show an increasing trend indicating the onset of dust formation. However, no corresponding
sharp decline in the optical light curve is seen at this stage implying that any dust, that
may have formed, is optically thin.
Such kind of dust formation was observed earlier in few novae, for example V1668 Cyg (Gehrz et al. 1980),
Nova Aql 1995 (Mason et al. 1996; Kamath et al. 1997) and  Nova LMC 1988 $\#$1 (Schwarz et al. 1998).
In V1668 Cyg dust formed in between days 30 - 60 after outburst and reached a visual optical depth $\sim$ 0.1 around 60 days after outburst; in Nova Aql 1995 dust formed shortly (about 9 days) after discovery and
in Nova LMC 1988 $\#$1 the presence of dust which lasted for about 2 months, was detected $\sim$ 55 days after visual maximum.
In all these novae no "dip" in the optical light curve was observed which is generally observed
in the visual light curve of
optically thick dust forming novae, for example V1280 Sco (Das et al. 2008), V705 Cas (Evans et al. 1997),
as the optical photons are absorbed by the dust shell.
The reason for finding optically thin dust is probably the clumpiness of the dust shell
so that the most of the shorter-wavelength radiation passes through the gaps.
 Verification of such clumpiness may be sought by expecting  more structures in the line profiles; structures caused by individual clumps having different brightness and kinematic velocities. However  the spectral resolution of the present data of $\sim$ 1000 is too low to discern or to be certain of the presence of such structures.
An alternate reason for optically thin dust could be formation of less number of grains which is not sufficient
to cover the entire solid angle along the line of sight of the observer. Though the observational data is unable to point out the exact reason for such optically thin dust formation, sometimes the clumpiness can be inferred from visual observations of old novae (Gehrz 1988).

We have attributed the steady brightening  of the $J,H,K$  fluxes  after October 15 to the onset of dust formation. The increase in the NIR fluxes being due to an increase in emission line strengths is also a possibility but extremely unlikely for the following reasons. Using IRAF, we estimate that the combined line flux of all lines in a band is typically  10 percent of the total broad-band flux in that band. Thus for example, one can consider what would happen if the $K$  band brightening by $\sim$ 2 magnitudes ( corresponding to a  flux increase by a very large factor factor of 6.25) between 15 October  and 3 November were  completely due to $K$  band lines (primarily Br $\gamma$) becoming stronger while the continuum strength remained constant (i.e. no contribution from dust formation to the continuum took place). In such a case the lines would have to necessarily become extremely strong and the signal at the line peak, compared to the continuum value,  would have to become much stronger that earlier (we estimate a factor of 5 to 10 increase in the peak-to-continuum ratio for most lines). Such behavior is just not seen in the present data. On the other hand, as we discuss in the next paragraph, the observed increase in the IR fluxes are very typical of the behavior associated with the onset of dust formation.

In case of V476 Sct we have studied the dust phase by analyzing  the spectral energy
distribution (SED) covering the $JHK$ bands in near-IR region. The near-IR fluxes, corresponding to the observed $JHK$ magnitudes after correction for extinction, are shown for 6 representative epochs starting from 8 Nov 2005 when the object had peaked in its $K$ band brightness. The SED's have been fitted with black body curves, the corresponding temperatures of which are indicated in Figure 5.  As may be seen the formal black-body temperatures of the dust shell gradually decrease with time  starting from 1615 K on 8 November 2005 to 1040 K on the last day of the observations on 2 December 2005. The shift in the peak of the SED with time towards longer wavelengths is clearly evident and is suggestive of the dust cooling with time. The black body temperatures may not represent the actual dust temperatures in an accurate manner for two reasons. First, the data covers only the 1 to 2.5 $\mu$m region and the tail of the distribution (the Rayleigh-Jeans regime) is not sampled. Second, the emissivity of actual dust grains depends on its composition and actual grain size distribution and the  frequency dependence of the emissivity  can deviate from that of a blackbody (Krugel 2003). However our blackbody estimates are consistent with similar estimates of dust temperatures of $\sim$ 1000 K in case of V1668 Sgr,  1000 - 1200 K in Nova Aql 1995 and 950 K in LMC 1988 $\#$ 1.

The dust that formed towards the end of  2005 in nova V476 Sct persisted for quite some more time since the source was also detected by the AKARI infrared astronomical satellite which observed the whole sky in far IR (50-180 $\mu$m) and mid-IR (9 and 18$\mu$m) between May 2006 and August 2007. The source was detected in 3 and 2 passes  respectively during this period (AKARI does not list exact detection dates) in the 18 and 9 $\mu$m bands respectively with a mean flux of F(9$\mu$m) =  0.3138 $\pm$ 0.149 Jy and F(18$\mu$m) =  0.3919 $\pm$ 0.0911 Jy (equivalently 1.18x10$^{-18}$ W cm$^{-2}$ $\mu$m$^{-1}$ and 0.36x10$^{-18}$ W cm$^{-2}$ $\mu$m$^{-1}$). However by March-April of 2010, observations from the Wide field Infrared Survey Explorer (WISE) did not detect the source in any of the  3.4 (W1), 4.6 (W2), 12 (W3) and 22 $\mu$m (W4) bands indicating
that either the dust had been destroyed by radiation from the central white dwarf or its emission had fallen below the WISE detection limits.

\section{Summary}
The near-IR $JHK$ spectroscopic and photometric results of the dust forming nova V476 Sct
(2005) are presented. $JHK$ spectra and light curve are shown and spectral lines are identified.
From the IR spectra, the nova is classified to be of the FeII type consistent with the classification
based on optical spectra.
The key result concerns the finding of formation of optically thin dust shell in
the nova. Reported instances of novae which form such dust shells are few
and they are characterized by not displaying  a deep or significant dip in their
optical lightcurves at the onset of dust formation. The SEDs on several epochs,
since the date of commencement of dust formation are plotted,  allowing estimates of the
dust shell temperature to be made and also showing the gradual cooling of the dust with time.

\section{Acknowledgements}The research work at the Physical Research Laboratory is funded by the Department of Space, Government of India. The research work at the S N Bose National Centre for Basic Sciences is funded by Department of Space and Technology, Government of India. We acknowledge the variable star observations from the American Association of Variable Star Observers (AAVSO) and Association Francaise des Observateurs d$\prime$Etoiles Variables (AFOEV) data base, operated at CDS, France for the use of their optical photometric data contributed by several observers worldwide.

\end{document}